\documentclass[aps,prb,twocolumn, amsmath, amssymb,showpacs,superscriptaddress]{revtex4-1}

\usepackage{graphicx}
\usepackage{amsmath,amssymb,amsfonts}
\usepackage{textcomp}
\usepackage{hyperref}
\usepackage{gensymb}
\usepackage{verbatim}
\usepackage{bm}
\usepackage{amsmath}
\hypersetup{breaklinks=true,colorlinks=true,urlcolor=black}
\usepackage{color}
\usepackage{soul}
\DeclareGraphicsExtensions{.jpg,.pdf,.png,.eps}

\begin{document}

\title{Single crystal growth and characterization of antiferromagnetically ordering EuIn$_2$}
\author{Brinda Kuthanazhi}
\affiliation{Ames National Laboratory, Iowa State University, Ames, Iowa-50011, USA}
\affiliation{Department of Physics and Astronomy, Iowa State University, Ames, Iowa-50011, USA}
\author{Simon X.~M.~Riberolles}
\affiliation{Ames National Laboratory, Iowa State University, Ames, Iowa-50011, USA}
\affiliation{Department of Physics and Astronomy, Iowa State University, Ames, Iowa-50011, USA}
\author{Dominic H. Ryan}
\affiliation{Department of Physics, McGill University, Montreal, Québec H3A 2T8, Canada}
\author{Philip~J.~Ryan}
\affiliation{Advanced Photon Source, Argonne National Laboratory, Argonne, Illinois 60439, USA}
\author{Jong-Woo Kim}
\affiliation{Advanced Photon Source, Argonne National Laboratory, Argonne, Illinois 60439, USA}
\author{Lin-Lin Wang}
\affiliation{Ames National Laboratory, Iowa State University, Ames, Iowa-50011, USA}
\author{Robert J. McQueeney}
\affiliation{Ames National Laboratory, Iowa State University, Ames, Iowa-50011, USA}
\affiliation{Department of Physics and Astronomy, Iowa State University, Ames, Iowa-50011, USA}
\author{Benjamin G. Ueland}
%\email{bgueland@ameslab.gov}
\affiliation{Ames National Laboratory, Iowa State University, Ames, Iowa-50011, USA}
\affiliation{Department of Physics and Astronomy, Iowa State University, Ames, Iowa-50011, USA}
\author{Paul C. Canfield}
\email{canfield@ameslab.gov}
\affiliation{Ames National Laboratory, Iowa State University, Ames, Iowa-50011, USA}
\affiliation{Department of Physics and Astronomy, Iowa State University, Ames, Iowa-50011, USA}

\date{\today}

\begin{abstract}
We report the single crystal growth and characterization of EuIn$_2$, a magnetic topological semimetal candidate according to our density functional theory (DFT) calculations. We present results from electrical resistance, magnetization, M\"ossbauer spectroscopy, and X-ray resonant magnetic scattering (XRMS) measurements. We observe three magnetic transitions at $T_{\text{N}1}\sim 14.2~$K, $T_{\text{N}2}\sim12.8~$K and $T_{\text{N}3}\sim 11~$K, signatures of which are consistently seen in anisotropic temperature dependent magnetic susceptibility and electrical resistance data. M\"ossbauer spectroscopy measurements on ground crystals suggest an incommensurate sinusoidally modulated magnetic structure below the transition at $T_{\text{N}1}\sim 14~$K, followed by the appearance of higher harmonics in the modulation on further cooling roughly below $T_{\text{N}2}\sim13~$K, before the moment distribution squaring up below the lowest transition around $T_{\text{N}3}\sim 11~$K. XRMS measurements showed the appearance of magnetic Bragg peaks below $T_{\text{N}1}\sim14~$K, with a propagation vector of  $\bm{\tau}$ $=(\tau_h,\bar{\tau}_h,0)$, with $\tau_h$varying with temperature, and showing a jump at $T_{\text{N}3}\sim11$~K. The temperature dependence of $\tau_h$  between $\sim11$~K and $14$~K  shows incommensurate values consistent with the M\"{o}ssbauer data. XRMS data indicate that $\tau_h$ remains incommensurate at low temperatures and locks into $\tau_h=0.3443(1)$.  
\end{abstract}

\maketitle

\section{Introduction}
The search for new topological materials and understanding their physics has driven new materials physics research in the last few years. \cite{Hsieh2008, Hasan2010, Burkov2011, Burkov2016, Chiu2016, Yan2017, Armitage2018, Tokura2019} The robust surface states and the intriguing magneto-transport and optical behaviors of topological materials, such as quantum anomalous Hall effect, axion behavior, circular polarized galvanic effect, etc. \cite{Wan2011, Parameswaran2014, Chan2016, Jia2016, Chan2017, Shekhar2018} are expected to further our understanding of quantum materials as well as drive technology innovations. \cite{Yan2017,zhang2019} 

One such interesting class of materials is Weyl semimetals (WSM), which can be attained by breaking either the time reversal symmetry (TRS) or the inversion symmetry of a parent Dirac semimetal, which has degenerate bands crossing with a linear dispersion, hence hosting massless electronic excitations. In magnetic Weyl semimetals, magnetism coexists with and modifies the non-trivial topology of the electronic bands of the material, by breaking symmetries, through either providing a wave vector associated with a low temperature antiferromagnetic (AFM) state, or by imposing a long-range internal field via adopting a state with an ferromagnetic (FM) component. In case of FM-WSM, magnetism breaks time reversal symmetry, and in AFM-WSM, an effective TRS, which could be a combination of TRS and a discrete crystal symmetry such as translation.

Divalent Eu-containing binary and ternary compounds have become fertile grounds for looking for magnetic Weyl semimetal candidates. \cite{May2014, Soh2019C, Jo2020, Pakhira2020, Riberolles2021, LlWang2021} When present in its $2+$ valence, Europium is magnetic with an effective angular momentum $J=S+L=7/2$ that is fully associated with the electron spin $S = 7/2$ with $L$ being zero. In many cases, magnetism can be tuned by chemical doping \cite{Jo2020}, hydrostatic pressure \cite{Gati2021}, strain, etc., which then leads to many possibilities of manipulating topological phases in these systems. \cite{LlWang2021}

Recently a theoretical study predicted that EuTl$_2$, which forms in the CaIn$_2$ structure type with the hexagonal structure in space group $P6_3mm/c$ (194), to host a variety of topological phases depending on the strength of the magnetic interactions, as well as by varying strain. \cite{LlWang2021} EuIn$_2$, which is much more amenable to growths than the toxic and volatile Tl, crystallizes in the same CaIn$_2$ structure and could be a possible candidate for various topological phases. In the literature, EuIn$_2$ was synthesized in polycrystalline form and magnetic measurements were only done for temperatures above $\sim100$~K. \cite{Iandelli1964, Yatsenko1983} However, no reports exist about the low-temperature behavior of this compound and, therefore, the nature or even existence of magnetic order is unknown. As such EuIn$_2$ is a compound that merits further investigation in order to clarify its status as a potentially topologically non-trivial material.

Here, we report the results of DFT calculations on EuIn$_2$, predicting the existence of Dirac points in the non-magnetic state, which makes it a candidate for a magnetic topological semimetal if magnetic ordering can be confirmed and delineated. We report the single crystal synthesis, electrical transport measurements, and magnetic characterization of EuIn$_2$. We find signatures of three magnetic ordering temperatures at $T_{\text{N}1}\sim 14.2~$K, $T_{\text{N}2} \sim 12.8~$K and $T_{\text{N}3}\sim 11~$K in temperature dependent magnetization and electrical resistance measurements. Further measurements of M\"{o}ssbauer spectroscopy and X-ray resonant magnetic scattering results provide initial insight into the nature of the magnetic ordering occurring below these temperatures. Taken together our results indicate that EuIn$_2$ is a promising candidate for further studies of magnetic and potential topological properties. 

\begin{figure}
	\centering
	\includegraphics[width=\linewidth]{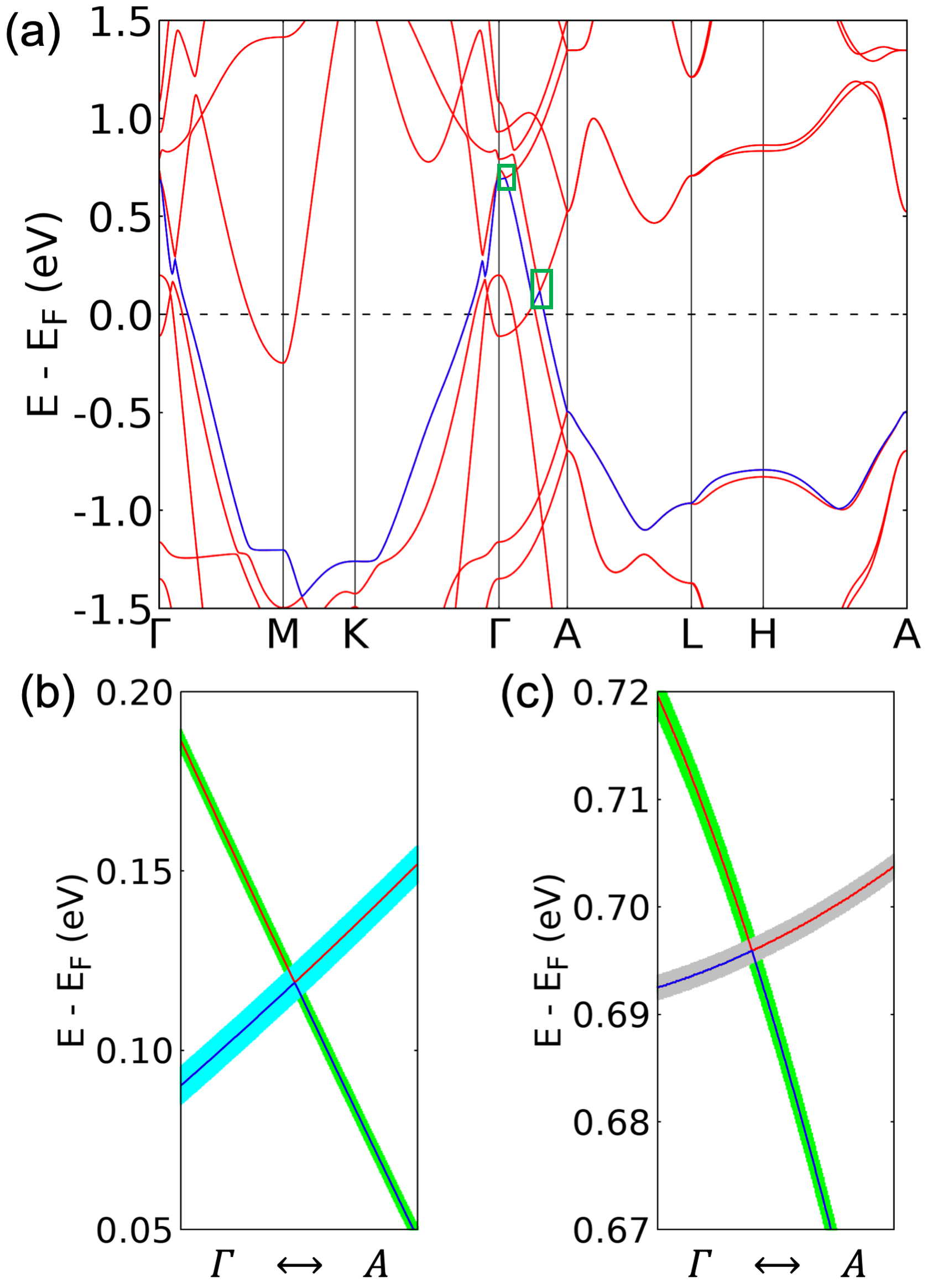}
	\caption{(a) Bulk band structure of non-magnetically-ordered EuIn$_2$ calculated in PBE+SOC without Eu 4f orbitals. Green rectangles mark the region that is shown in (b) and (c), zooming in near the two Dirac points along the $\Gamma-A$ direction. The top valence band is in blue. The green, cyan and grey shades stand for the projection of In $p_y$, In $s$ and Eu $d_{yz}$ orbitals, respectively.}
	\label{dft}
\end{figure}

\section{Experimental and Computational Methods}
\begin{figure*}
	\centering
	\includegraphics[width=\linewidth]{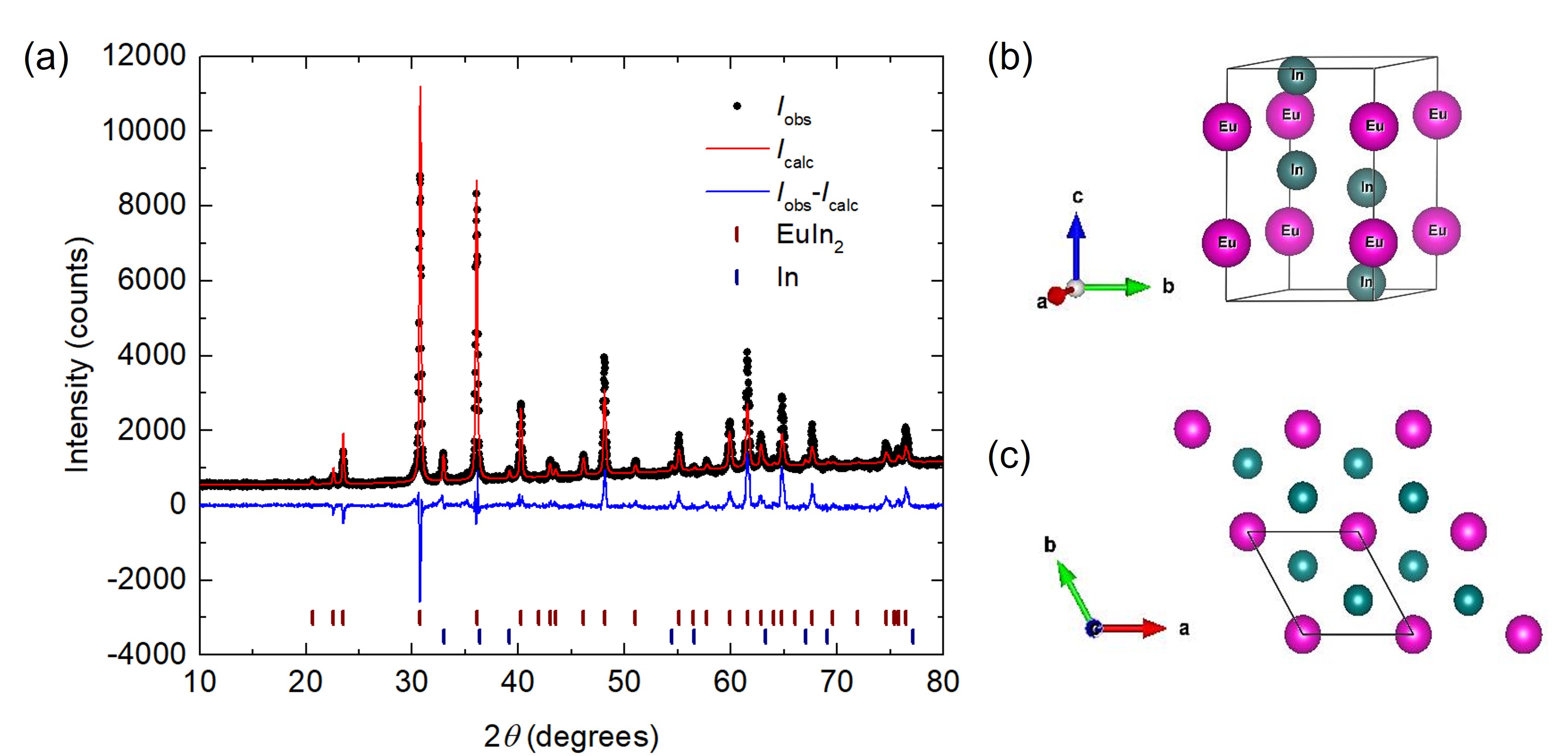}
	\caption{(a) Powder x-ray diffraction pattern with Rietveld refinement of EuIn$_2$, with lattice parameters $a=4.9788(7)~$\AA\ and $c=7.8667(5)~$\AA. The black line shows the measured pattern, the red line shows the calculated pattern and the blue line shows the difference between the two for EuIn$_2$. The red and blue bars show the Bragg peak positions for EuIn$_2$ and In, respectively. (b) Crystal structure of EuIn$_2$. (c) Projection of the structure along the crystallographic $c-$axis, showing the triangular lattice of Eu atoms.}
	\label{EuIn_xrd}
\end{figure*}

\begin{figure*}
	\centering
	\includegraphics[width=\linewidth]{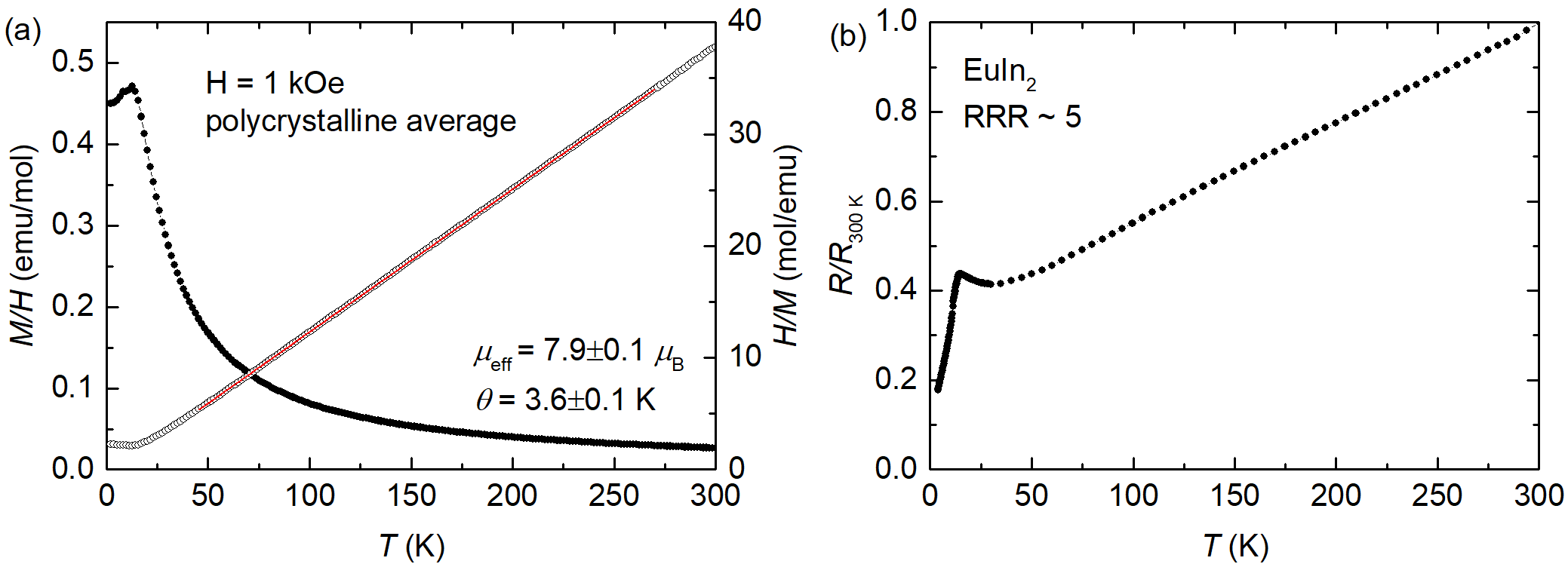}
	\caption{(a) Polycrystalline average of the magnetic susceptibility $M(T)/H$ along with the right $y-$axis showing the inverse susceptibility $H/M(T)$, with the linear fit showing the Curie-Weiss behavior. (b) Normalized electrical resistance $R/R_{300 K}$ varying with temperature. All data below $T=3.5~$K is removed as In flux present becomes superconducting below this temperature.}
	\label{MTRT}
\end{figure*}

Single crystals of EuIn$_2$ were grown out of excess In using the high-temperature solution growth method. \cite{Canfield1992, Canfield2016, Canfield2019} Elements (Eu - Ames Laboratory, 99.99+ \% and In - Alfa Aesar, 99.999\%) with an initial stoichiometry of Eu$_{15}$In$_{85}$ were weighed out into a fritted alumina crucible set (Canfield Crucible Set) \cite{Canfield2016} and sealed in a fused silica tube with a partial pressure of Argon. The prepared ampoule was then heated up to $800~^\circ$C over 4 hours and held there for 3 hours. This was followed by a slow cooling to $480~^\circ$C over 50 hours and decanting the excess flux using a centrifuge. \cite{Canfield1992, Canfield2019} The choice of decanting temperature at $480~^\circ$C is to avoid the possible formation of EuIn$_4$ below the $450~^\circ$C peritectic line (ASM Diagram \#901007 \cite{OkamotoEuIn}). The crystals obtained had a hexagonal morphology and were air sensitive, with the surface turning white/oxidized within about 15 minutes of exposure to air. They were stored and handled inside a Nitrogen filled glovebox. 

Powder x-ray diffraction measurements were carried out to determine the phase purity. The powder x-ray diffraction pattern was collected on ground crystals using a Rigaku Miniflex diffractometer inside a Nitrogen filled glovebox, with Cu$K_{\alpha}$ radiation. Crystallographic parameters were obtained from a Rietveld refinement using GSAS-II software. \cite{Toby2013} 

The magnetic measurements were carried out in a Quantum Design Magnetic Property Measurement System with applied fields up to $70~$kOe. The samples were mounted on a Poly-Chloro-Tri-Fluoro-Ethylene disk, and the separately measured background of the disk was subtracted. The AC resistivity measurements were done in a Quantum Design Physical Property Measurement System in the standard four point configuration. The current was passed along the $c-$axis with $I=3~$mA and $f=17~$Hz. Contacts were made using Dupont-4929N silver paint.

The $^{151}$Eu M\"ossbauer spectroscopy measurements were carried out on ground samples of EuIn$_2$, using a 4~GBq $^{151}$SmF$_3$ source, driven in sine-mode and calibrated using a standard $^{57}$Co\underline{Rh}/$\alpha$-Fe foil.
Isomer shifts are quoted relative to EuF$_3$ at ambient temperature. The sample was hand-ground under hexane (to minimise oxidation) and mixed with boron nitride before being loaded into a thin-window delrin sample holder. The sample was cooled in a vibration-isolated closed-cycle helium refrigerator with the sample in helium exchange gas. The spectra taken below 10~K and above 15~K could be fitted conventionally using a  sum of Lorentzian lines with the positions and intensities derived from a full solution to the nuclear Hamiltonian. \cite{voyer} However, spectra taken in the incommensurate modulated phase between 10~K and 15~K (see discussions below) were fitted using a model that derives a distribution of hyperfine fields from an (assumed) incommensurate sinusoidally modulated magnetic structure. \cite{bonville349, maurya216001}

X-ray resonant magnetic scattering (XRMS) measurements were made at end station $6$-ID-B at the Advanced Photon Source, Argonne National Laboratory, for the Eu $L_2$ edge ($E=6.170$~keV \cite{Sole2007}) using the Huber Psi-circle geometry diffractometer. An approximately $0.5$~mm thick single-crystal with a cross section of $\approx 1~\text{mm}\times1$~mm was mounted on a Cu sample holder with carbon tape and the holder was thermally anchored to the cold head of a He displex with the $\mathbf{c}$ crystalline axis horizontal.  Care was taken to minimize the exposure time of the sample to air.  The instrument was operated in vertical geometry which allowed for access to several $(h\,k\,l)$ reciprocal-lattice points. Incident x-rays were linearly polarized perpendicular to the vertical scattering plane ($\sigma$ polarized) and no polarization analysis of the diffracted beam was performed.  A $2$-dimensional Pilatus 100K detector was employed and the incident beam size was $0.248~\text{mm}\times0.600$~mm (height by width). An attenuator with a calculated transmission of $0.396564$ was used when recording the data shown below in order to mitigate beam heating of the sample. Nevertheless, comparison of the XRMS data with the thermodynamic, transport, and M\"ossbauer spectroscopy data indicated a $T\approx1$~K difference in features in the data which is presumably due to beam heating. \textit{We have accordingly shifted the temperature of the XRMS data presented below by $1$~K.}

Band structures of non-magnetically-ordered EuIn$_2$ have been calculated without Eu 4f orbitals in density functional theory\cite{Kohn1964, Kohn1965} using PBE\cite{Perdew1996} as exchange-correlation functional with spin-orbit coupling (SOC) effect included. All DFT calculations have been performed in VASP\cite{Kresse1996, GKresse1996} with a plane-wave basis set and projector augmented wave\cite{Blochl1994} method. We used the hexagonal unit cell of 6 atoms with a $\Gamma$-centered Monkhorst-Pack\cite{Pack1976} ($10\times 10\times 6$) $k-$point mesh with a Gaussian smearing of 0.05~eV. The kinetic energy cutoff was 400~eV. 

\section{Results and Discussion}
\begin{figure}
	\centering
	\includegraphics[scale=1]{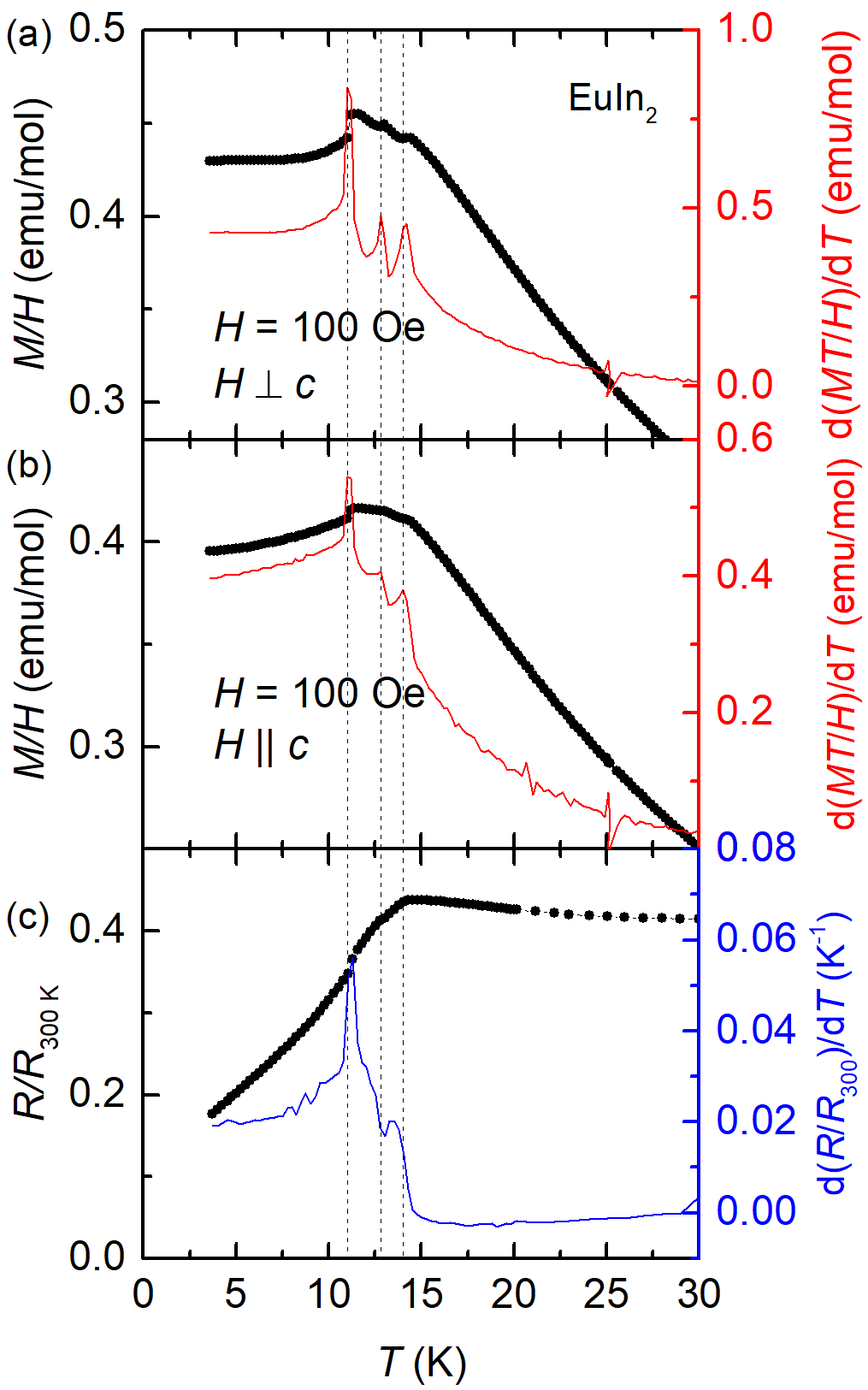}
	\caption[$M(T)/H$ and $R(T)$ with the temperature derivatives $d(MT/H)/dT$ and $dR/dT$ for EuIn$_2$]{Magnetic susceptibility $M(T)/H$ and the temperature derivative $d(MT/H)/dT$ as a function of $T$, for an applied field of $H=100~$Oe with (a)  $H\perp c$. (b) $H\parallel c$. (c) Normalized resistance $R/R_{300 K}$ and the temperature derivative of it $d(R/R_{300})/dT$. The data below $T=3.5~$K are removed as the In flux present becomes superconducting below this temperature. The three vertical dashed lines indicate the temperature values of local maxima in $d(MT/H)/dT$ and cross through $d(R/R_{300})/dT$ data near maximum local slope points.}
	\label{MTRTd}
\end{figure}
\subsection{DFT Results}
The bulk band structure of non-magnetically-ordered EuIn$_2$, without Eu 4f orbitals, is plotted in Fig. \ref{dft}(a) with the highest valence band in blue according to simple filling. The valence and conduction bands are generally well removed from the Fermi level, E$_\text{F}$, over most of the Brillouin zone except for an electron pocket at the $M$ point and the overlap between valence and conduction bands around the $\Gamma$ point. Especially along the $\Gamma-A$ direction, there are two gapless crossing points between the top valence and bottom conduction bands protected by the three-fold rotational symmetry and are Dirac points. As zoomed in Fig.\ref{dft}(b) and (c), these two Dirac points are above E$_\text{F}$ and have the momentum-energy of (0, 0, ±0.24 \AA$^{-1}$; E$_\text{F}$+0.12 eV) and (0, 0, ±0.04 \AA$^{-1}$; E$_\text{F}$+0.70~eV), respectively. The lower Dirac point has a switch of orbital character between In $p_y$ and In $s$, whereas the upper Dirac point is between In $p_y$ and Eu $d_{yz}$. Similar to EuTl$_2$\cite{LlWang2021}, these Dirac points can act as the parent gapless phase to be transformed into different descendent phases, depending on the existence and nature of magnetic ordering. As such, it is of interest to know what type of magnetic structures EuIn$_2$ hosts. 

\subsection{Powder X-ray Diffraction, Resistivity, and Magnetization}

\begin{figure*}
	\centering
	\includegraphics[width=\linewidth]{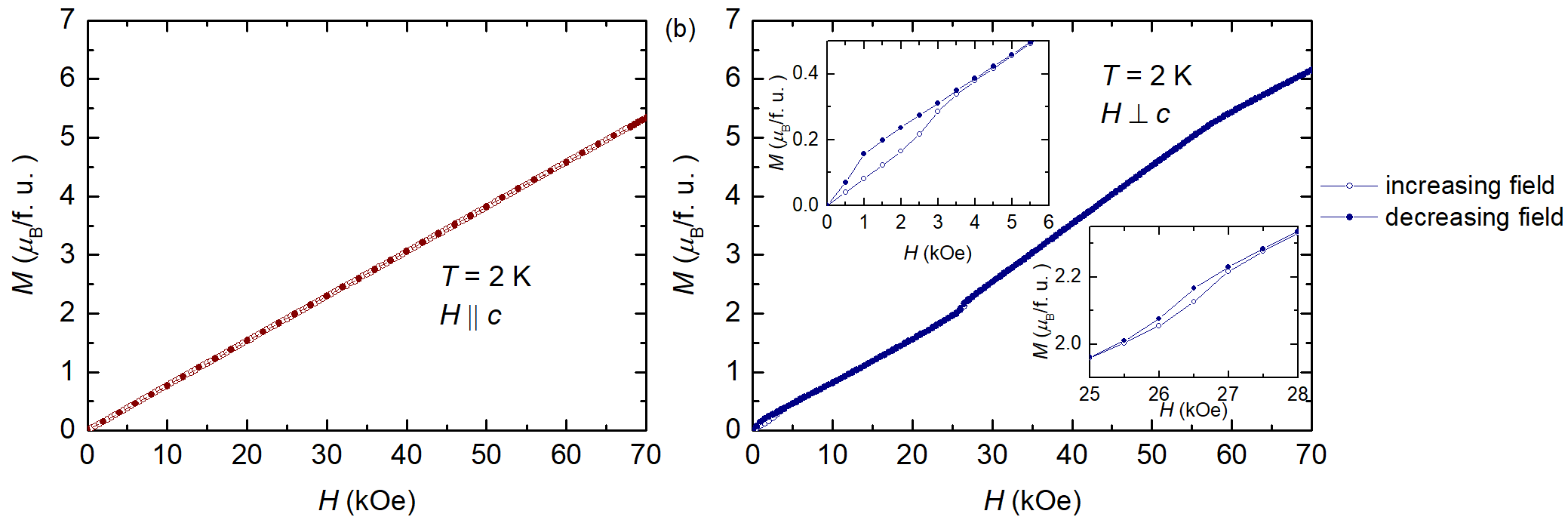}
	\caption[Magnetization $M(H)$ data for EuIn$_2$]{Magnetization as a function of applied field, $M(H)$, for fields up to $70~$kOe, measured at $T=2~$K, with the field directions (a) $H\parallel c$ and (b) $H \perp c$. Insets show the zoom-in of the field regions close to $H=1~$kOe and $H=26~$kOe, showing the hysteretic nature of the two transitions. The open and closed symbols denote increasing and decreasing field measurements, respectively.}
	\label{MH}
\end{figure*}
\begin{figure}
	\centering
	\includegraphics[scale=0.7]{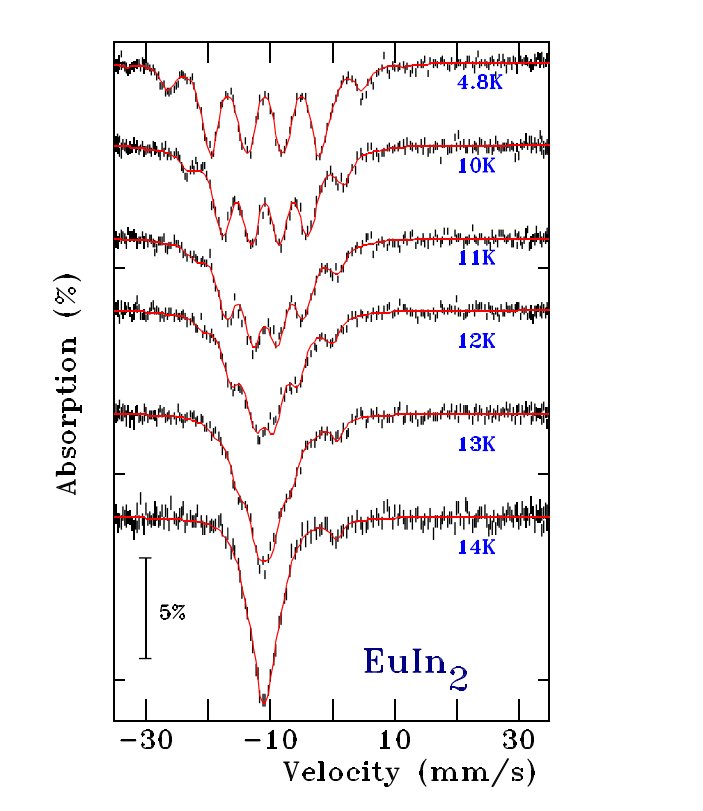}
	\caption[M\"ossbauer spectra at various temperatures for EuIn$_2$]{$^{151}$Eu M\"ossbauer spectra of EuIn$_2$ at various temperatures between 4.8~K to 14~K. The red solid lines are fit using the modulated model explained in the text.}
	\label{Mossbauer1}
\end{figure}

The powder x-ray diffraction pattern shown in Fig. \ref{EuIn_xrd}(a) confirms the $P6_3mm/c$ structure of EuIn$_2$, with lattice parameters $a=4.9788(7)~$\AA\ and $c=7.8667(5)~$\AA\ obtained from Rietveld refinement (with the goodness of fit $wR=9.98$\%). Some impurity peaks are present, which can be identified as In, with about 6\% phase fraction of In present. Given that the EuIn$_2$ is grown out of an excess of In, this is the expected impurity. The In most likely comes from small droplets of solidified, excess liquid that adhered to the EuIn$_2$ crystals through the decanting step. Lattice parameters obtained from Rietveld refinement agree with the existing report $a=4.9750~$\AA\ and $c=7.8690~$\AA. \cite{Iandelli1964} 
Figure \ref{EuIn_xrd}(b) shows the three dimensional crystal structure of EuIn$_2$, and (c) shows the projection along the $c-$axis. The Eu atoms form a triangular lattice in the $ab-$plane, a geometry conducive to magnetic frustration.

We first survey the temperature dependence of $M(T)$ and $R(T)$ over a wide temperature range ($300~$K $> T >3.5~$K). The low-field magnetization data and the resistivity data shown in Fig. \ref{MTRT} and Fig. \ref{MTRTd} are truncated at $T=3.5~$K as the In flux present becomes superconducting below this temperature which gives rise to small discontinuities. The polycrystalline average of the magnetic susceptibility was obtained using $(M/H)_{poly}=\frac{1}{3} (M/H)_{parallel}+\frac{2}{3}(M/H)_{perp}$, where the subscripts `parallel' and `perp' denote the applied field directions $H||c$ and $H\perp c$ respectively. Figure \ref{MTRT}(a) shows the polycrystalline average of $M/H$ plotted as a function of temperature, for an applied field of $H=1~$kOe. The high-temperature region shows a Curie-Weiss behavior. The right-hand axis of Fig. \ref{MTRT}(a) shows the inverse susceptibility with a linear fit. The fitting parameters of the linear fit gave an effective moment of $\mu_{eff} = 7.9\pm0.1 \mu_B$ (consistent with the 7.94~$\mu_B$ theoretically anticipated for Eu$^{2+}$) and $\theta = 3.6\pm0.1~$K. Normalized electrical resistance $R/R_{300 K}$ is plotted in Fig. \ref{MTRT}(b). The sample has a residual resistivity ratio of $\sim5$. The low-temperature resistivity shows a slight upturn followed by a sharp decrease due to loss of spin disorder scattering. A signature characteristic of antiferromagnetic ordering below $\sim15~$K is clearly seen in both $M(T)/H$ as well as $R(T)$ data.

\begin{figure*}
	\centering
	\includegraphics[scale=0.6]{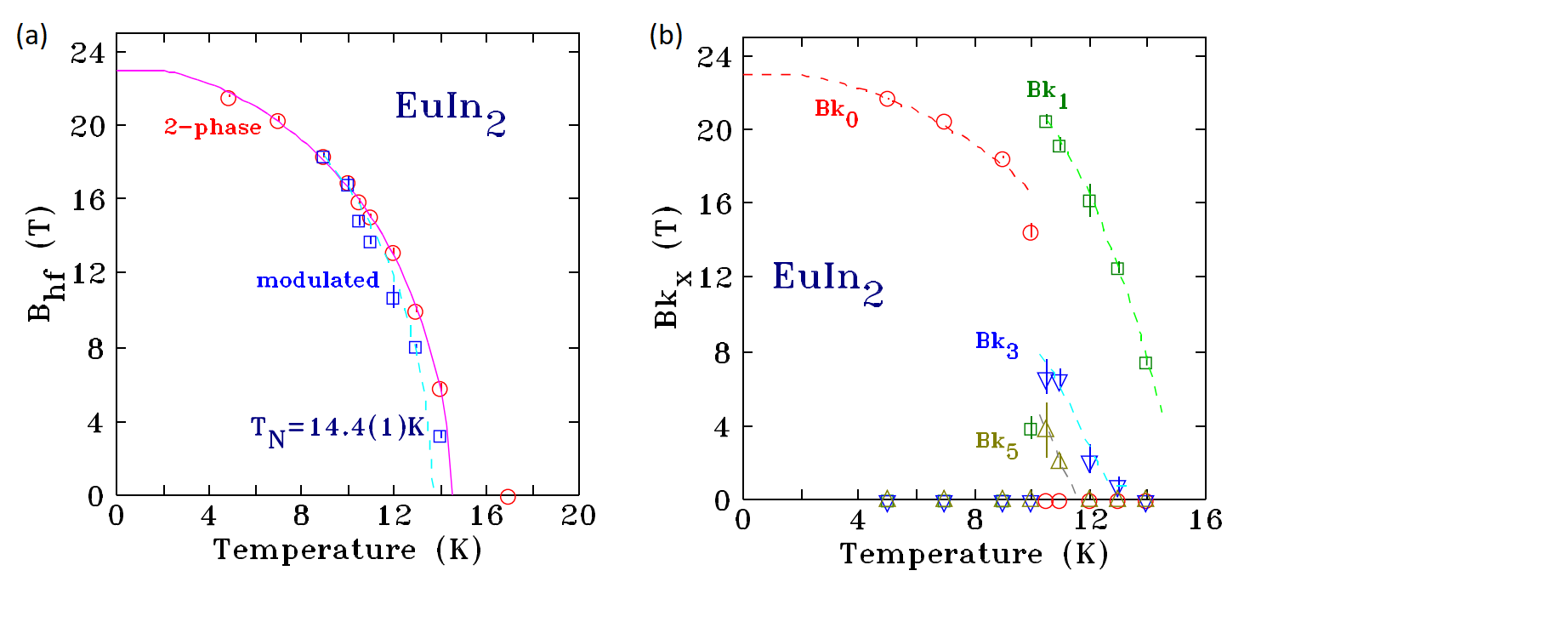}
	\caption[Hyperfine field varying with temperature from M\"ossbauer]{(a) Temperature dependence of the hyperfine field, $B_{hf}$, obtained from fitting the M\"ossbauer spectra using a two-component fit for EuIn$_2$ shown in red circles. The temperature dependence of the average of B$_k$ from the modulated fit is shown in blue square symbols. (b) Temperature variations of the Fourier components of the hyperfine field from fitting the M\"ossbauer spectra using a modulated model.}
	\label{Mossbauer2}
\end{figure*}

%Figure: XRMS data
\begin{figure}[h!]
	\centering
	\includegraphics[width=1\linewidth]{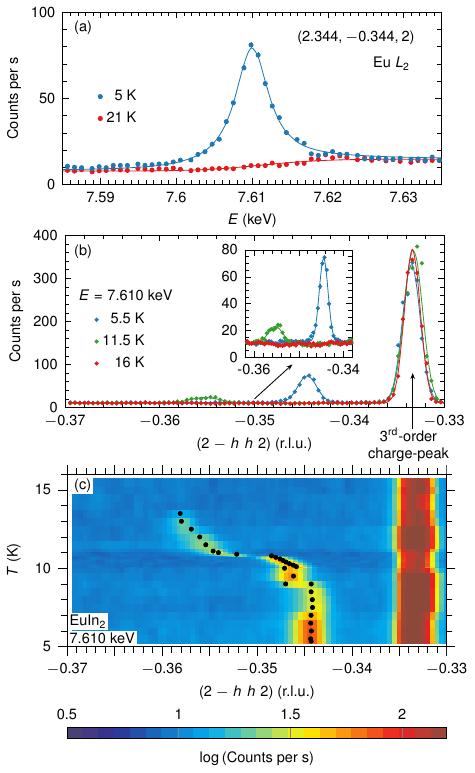}
	\caption{  \label{Fig:XRMS} (a)  X-ray energy scan across the Eu $L_{2}$ edge for $(2+\tau_h,\bar{\tau}_h,2)$ at $21~$K and $5$~K.  (b) Diffraction patterns across the $(2+\tau_h,\bar{\tau}_h,2)$ magnetic-Bragg peak using $7.610$~keV x-rays at $5.5~$K, $11.5~$K, and $16~$K. As explained in the text, the peak at $(2\frac{1}{3},-\frac{1}{3},2)$ is due to the $(7,\bar{1},6)$ structural-Bragg peak. (c) Image plot showing a detailed temperature dependence of the $(2+\tau_h\,\bar{\tau}_h\,2)$ magnetic-Bragg peak from $7.610$~keV XRMS data. Black dots mark the centers of the magnetic-Bragg peak as determined  from fits to gaussian lineshapes. Error bars are smaller than the symbol size if they are not visible. All temperatures presented are after shifting by $\approx1~$K, as discussed in the text.}
\end{figure}

As we go to lower temperatures, we see signatures of up to three transitions in the magnetization data between  $10-15~$K. These can be clearly seen in  the anisotropic magnetic susceptibilities and resistivity along with their derivatives, shown in Fig. \ref{MTRTd}. Figure \ref{MTRTd}(a) shows magnetic susceptibility $M/H$, measured under field cooling, and the temperature derivative $d(MT/H)/dT$ \cite{Fisher1962} as a function of $T$, for an applied field of $H=100~$Oe, with $H\perp c$. Similar data for $H\parallel c$ is shown in Fig. \ref{MTRTd}(b). Figure \ref{MTRTd}(c) shows the normalized resistance $R/R_{300 K}$ and its temperature derivative, $d(R/R_{300 K})/dT$, \cite{Fisher1968} on the right $y-$axis. The signatures of three transitions are seen in the derivatives of both $M(T)/H$ data sets, with transition temperatures of $T_{\text{N}1} = 14.2\pm0.2, T_{\text{N}2} = 12.8\pm0.2$, and $T_{\text{N}3} = 11.0\pm0.2~$K. Similar transition temperatures can be inferred from the $dR/dT$ data. Whereas the two higher temperature transitions have features consistent with second order phase transitions, the more discontinuous nature of the $M(T)$ and $R(T)$ data for the lowest temperature transition, along with the shape of the derivative curves, suggest that $T_{\text{N}3}\sim11~$K may be associated with a first order phase transition. 

Magnetization as a function of field $M(H)$ data for $T=2~$K is shown in Fig. \ref{MH}(a) for $H \parallel c$ and (b) for $H\perp c$. Superconducting Indium has a critical field of $H_c\sim250~$Oe, at $T\sim1~$K, \cite{Shaw1960} so the effects on $M(H)$ data are minimal. Also, In features, would be isotropic because the In is primarily in the form of tiny polycrystalline droplets or streaks adhering to the surface of the crystal. The measurement was done by cooling the sample to $2~$K in zero field and then measuring while increasing and decreasing the magnetic field. The $H\parallel c$ data show linear behavior, with increasing and decreasing fields overlapping each other. On the other hand, $M(H)$ for $H\perp c$ show subtle features corresponding to various possible spin re-orientation transitions. There is a low field, hysteretic transition between $1-3~$kOe, shown more clearly in the inset of Fig. \ref{MH}(b). There are two more subtle features in the $M(H)$; one around $27~$kOe also shows a small hysteresis [second inset of Fig. \ref{MH}(b)], and another change of slope around $57~$kOe where no discernible hysteresis is observed. 

\subsection{M\"ossbauer Spectroscopy}
%Figure: Scattering Parameters
\begin{figure}[h!]
	\centering
	\includegraphics[width=1\linewidth]{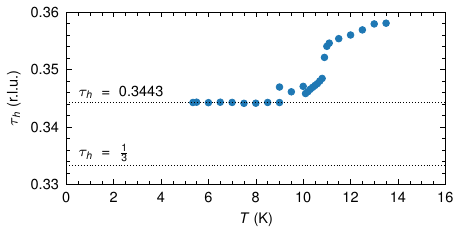}
	\caption{  \label{Fig:Scat_Params} The temperature dependence of $\tau_h$, the component of the three symmetry equivalent antiferromagnetic propagation vectors $\bm{\tau}=(\tau_h,0,0)$, $(0,\tau_h,0)$, and $(\bar{\tau}_h,\tau_h,0)$ determined from XRMS measurements of the $(2+\tau_h,\bar{\tau}_h,2)$ magnetic-Bragg peak of a single-crystal sample. The temperatures presented are after shifting by $\approx1~$K, as discussed in the text.} 
\end{figure}
Figure \ref{Mossbauer1} shows $^{151}$Eu M\"ossbauer spectra for EuIn$_2$ taken at representative, low temperatures. There is a clear signature of magnetic ordering as well as the evolution of splitting as the temperature is lowered through 14~K. At 4.8~K the $^{151}$Eu M\"ossbauer spectrum of EuIn$_2$ shows a well-split magnetic pattern with a hyperfine field (B$_{hf}$) of 21.5(1)~T and an isomer shift of $-$10.74(3)~mm/s, values typical for a magnetically ordered Eu$^{2+}$ compound \cite{long1989}(Fig. \ref{Mossbauer1} top spectrum). No information on the ordering direction within the crystallographic cell can be obtained from this spectrum as the observed quadrupole contribution, 0.0(3)~mm/s is consistent with zero. On warming, the spectra show the expected narrowing as B$_{hf}$ decreases, but the spectral shape also changes, developing more weight in the center. A simple two-component fit (one magnetic with a temperature dependent B$_{hf}$, the other with no magnetic splitting) models the rapidly growing central area but does not yield satisfactory fits. However, fitting the derived B$_{hf}$(T) using a J=$\frac{7}{2}$ Brillouin function provides an estimated ordering temperature of 14.4(1)~K, consistent with $T_{\text{N}1}$, as shown in Fig. \ref{Mossbauer2}(a). We note that a $\sim$5\% Eu$^{3+}$ impurity is apparent in the higher temperature spectra (the weak line at $\sim$0~mm/s), which could be coming from some of the sample being oxidized, and this was included in all of the fits.

In order to fit the data with the modulated model, if we assume that the moment modulation along the direction of the propagation vector {\bf k} can be written in terms of its Fourier components, and further assume that the observed hyperfine field is a linear function of the Eu moment at any given site, then the variation of B$_{hf}$ with distance $x$ along the propagation vector {\bf k} can be written as \cite{bonville349}:
\begin{equation}
	B_{hf}(kx) = Bk_0 + \sum^n_{l=0} Bk_{2l+1} \sin(2l+1)kx
	\label{eqn:fourier}
\end{equation}
where the $Bk_n$ are the odd Fourier coefficients of the field modulation. As $+$B$_{hf}$ and $-$B$_{hf}$ are indistinguishable, $kx$ only needs to run over half the modulation period, and in this case, a square-wave modulated structure can be modeled either as a sum over a very large number of Fourier coefficients or by simply using the $Bk_0$ term with all of the other $Bk_n$ set to zero. We found the fits to be far more stable with the $Bk_0$ term included rather than using a large set of $Bk_n$, however, the two approaches are
effectively equivalent. Variations of this model have also been used to fit spectra of EuPdSb \cite{bonville349} and Eu$_4$PdMg. \cite{ryan17d108}

Adopting the incommensurate modulated model to analyze the spectra yields the fits shown in Fig. \ref{Mossbauer1}, and the temperature dependence of the derived modulation harmonics is shown in Fig. \ref{Mossbauer2}(b). Starting from the lowest temperature, we see that only Bk$_0$ is present in the 4.8~K spectrum. This suggests that the ground state is a squared-up state without moment modulations. On warming above $T_{\text{N}3}$, the higher harmonics Bk$_5$, Bk$_3$, and Bk$_1$ appears, indicating a modulated structure. On further warming, the higher harmonic contributions reduce, and only Bk$_1$ survives, indicating that the order evolves towards a purely sinusoidally modulated state before the order disappears at $\sim14~$K. 
 
Thus, the M\"ossbauer results confirm that: $T_{\text{N}1}$ at $\sim14~$K is a transition to an incommensurate antiferromagnetic state. $T_{\text{N}2}$ at $\sim13~$K might be associated with the start of the process of higher harmonics developing in the modulated order, and $T_{\text{N}3}$ at $\sim11~$K marks the completion of the squaring up of the moment modulation leading to all of the europium moments being equal. Further insight into the microscopic details of the Eu ordering requires diffraction measurements.

\subsection{X-ray Resonant Magnetic Scattering}

We used XRMS to determine the magnetic propagation vectors associated with the multiple magnetic phases below 14 K. Data from the scans described below determined the existence of an antiferromagnetic propagation vector of $\bm{\tau}=(\tau_h,\bar{\tau}_h,0)$. As explained earlier, all XRMS data discussed below are presented with temperature shifted by $\approx1~$K, to account for the beam heating. This makes the features seen in XRMS coincide well with those from M\"ossbauer spectroscopy, magnetization and resistance measurements. 

X-ray energy scans across the Eu $L_2$ edge were taken at $T=5$~K and $21$~K after aligning to the $(2+\tau_h\,\bar{\tau}_h\,2)$ magnetic-Bragg peak of our single-crystal sample at $5$~K and are shown in Fig.~\ref{Fig:XRMS}(a). A large resonant enhancement is seen just below the absorption edge around $E=7.610$~keV for $5$~K but not for $21$~K. This is consistent with an enhancement of dipole transitions of $2p$ core-level electrons to empty $5d$ states due to the presence of magnetic order at $5$~K. Based on these data, we made a series of longitudinal, rocking, and  other reciprocal-space scans using $7.610$~keV x-rays to characterize the temperature dependence of the magnetic-Bragg peaks. Figures~\ref{Fig:XRMS}(b) and (c) summarize the main results.

Figure~\ref{Fig:XRMS}(b) shows data from scans along $(2-h,h,2)$ at different temperatures. The $(2+\tau_h,\bar{\tau}_h,2)$ magnetic-Bragg peak is visible for $T=5.5$~K and $11.5$~K, but is absent for $16$~K. Fits to gaussian lineshapes find that the magnetic-Bragg peaks are centered at $h=-0.3443(1)$~r.l.u.\ and $h=-0.3554(1)$~r.l.u.\ for $5.5$ and $11.5$~K, respectively. The much stronger peak appearing at $h=-\frac{1}{3}$~r.l.u.\ in all three datasets is from the $(7,\bar{1},6)$  structural (charge) Bragg peak and arises from diffraction of x-rays with $\frac{1}{3}$ the wavelength of those corresponding to $E=7.610$~keV. This was verified by observing its negligble response to the insertion of x-ray attenuators which have a much greater effect on $7.610$~keV x-rays than on $22.830$~keV x-rays.

A more detailed temperature dependence of the \mbox{$(2+\tau_h,\bar{\tau}_h,2)$} magnetic-Bragg peak is given in Fig. \ref{Fig:XRMS}(c). The sample was realigned at each temperature before making a $(2-h,h,2)$ scan. Fits using a gaussian lineshape  find that the full-width at half maximum of the $(2+\tau_h,\bar{\tau}_h,2)$ peak does not change with decreasing temperature but that the center moves towards $(2\,0\,2)$.  This is shown by the black dots in Fig.~\ref{Fig:XRMS}(c) and the plot in Fig.~\ref{Fig:Scat_Params}. The magnetic-Bragg peaks appear at $T_{\text{N}1}\approx14$~K and there is a jump in $\tau_h$ at $T_{\text{N}3}\approx11$~K, which is further evidence for the first order nature of the transition at $T_{\text{N}3}$, which was suggested by the derivatives of $R/R_{300 K}$ and $MT/H$ data in Fig. \ref{MTRTd}. The temperature dependence of $\tau$  between $\approx11$~K and $14$~K  and its incommensurate value are consistent with the analysis of the M\"{o}ssbauer data. The XRMS data indicate that $\tau_h$ locks into $\tau_h=0.3443(1)$ at low temperature which is not an obviously commensurate value. Data in Figs.~\ref{Fig:XRMS}(b) and (c) show that the center of the third-order $(7,\bar{1},6)$ charge-Bragg peak does not change between $5.5$ and $16$~K which gives an excellent reference for the precision and accuracy we claim for the XRMS $\tau_h(T)$ data. Future neutron scattering experiments are needed to determine the details of the magnetic structure of EuIn$_2$.

\section{Conclusion}
In summary, we have synthesized single crystals of EuIn$_2$, which is a magnetic topological semimetal candidate according to DFT calculations. EuIn$_2$ undergoes three magnetic transitions with decreasing temperatures, between $10-15~$K. Clear signatures of the transitions are observed both in magnetic susceptibility and electrical resistivity measurements. Furthermore, M\"ossbauer spectroscopy measurements suggest a squared-up, likely incommensurate ground state, evolving into a complicated modulated moment order on warming, which eventually turns into a sinusoidally modulated order before turning paramagnetic above $T_{\text{N}1}$. XRMS data indicates an antiferromagnetic ordering with an incommensurate propogation vector, which changes with decreasing temperatures below $T_{\text{N}1}$, before locking in at $T_{\text{N}3}$, through a first order transition-like jump. It will be interesting to further explore the field-temperature behavior of the various transitions and determine the magnetic structure in various phases. Exploring the effects of various magnetic transitions on electronic band topology for this compound could be promising. 

\begin{acknowledgements}
Work at the Ames National Laboratory was supported by the U.S. Department of Energy, Office of Science, Basic Energy Sciences, Materials Sciences and Engineering Division. The Ames National Laboratory is operated for the U.S. Department of Energy by Iowa State University under Contract No. DEAC0207CH11358. B.K., P.C.C., S.X.M.R., L.-L.W, B.G.U., and R.J.M. were supported by the Center for the Advancement of Topological Semimetals, an Energy Frontier Research Center funded by the U.S. DOE, Office of Basic Energy Sciences. A portion of this research used resources at the Advanced Photon Source, which is a U.S. DOE SC User Facility operated by Argonne National Laboratory under Contract No.~DE-AC02-06CH11357. Financial support for this work was provided by Fonds Qu\'eb\'ecois de la Recherche sur la Nature et les Technologies, and the Natural Sciences and Engineering Research Council (NSERC) Canada. Some of this work was carried out, purely in English, while D.H.R. was on sabbatical at Iowa State University and their generous support during this visit is gratefully acknowledged.
\end{acknowledgements}

%\section*{Appendix}

\clearpage

\section*{References}

\bibliographystyle{apsrev}

\bibliography{EuIn2}

\end{document}